\documentclass[lettersize,journal]{IEEEtran}
\usepackage{amsmath,amsfonts}
\usepackage{algorithmic}
\usepackage{algorithm}
\usepackage{array}
\usepackage[caption=false,font=normalsize,labelfont=sf,textfont=sf]{subfig}
\usepackage{textcomp}
\usepackage{stfloats}
\usepackage{url}
\usepackage{verbatim}
\usepackage{graphicx}
\usepackage{cite}
\usepackage[utf8]{inputenc}
\begin{document}

\title{Hybrid Artificial-Living Cell Collectives for Wetware Computing}

\author{Ceylin Savas, Maryam Javed, Murat Kuscu
}

\author{Ceylin Savas, Maryam Javed,
        and Murat Kuscu,~\IEEEmembership{Member,~IEEE}
       \thanks{The authors are with the Nano/Bio/Physical Information and Communications Laboratory (CALICO Lab), Department of Electrical and Electronics Engineering, Koç University, Istanbul, Turkey. \emph{Corresponding author: Murat Kuscu (e-mail: mkuscu@ku.edu.tr)}.}
	   \thanks{This work was supported by The Scientific and Technological Research Council of Turkey (TUBITAK) under Grants \#123E516 and \#123C592.}}%


\maketitle

\begin{abstract}
Living systems continuously sense, integrate, and act on chemical information using multiscale biochemical networks whose dynamics are inherently nonlinear, adaptive, and energy-efficient. Yet, most attempts to harness such ``wetware'' for external computational tasks have centered on neural tissue and electrical interfaces, leaving the information-processing potential of non-neural collectives comparatively underexplored. In this letter, we study a hybrid artificial--living cell network in which \emph{programmable artificial cells} write time-varying inputs into a biochemical microenvironment, while a \emph{living bacterial collective} provides the nonlinear spatiotemporal dynamics required for temporal information processing. Specifically, artificial cells transduce an external input sequence into the controlled secretion of attractant and repellent molecules, thereby modulating the ``local biochemical context" that bacteria naturally sense and respond to. The resulting collective bacterial dynamics, together with the evolving molecular fields, form a high-dimensional reservoir state that is sampled coarsely (voxel-wise) and mapped to outputs through a trained linear readout within a physical reservoir computing framework. Using an agent-based \textit{in silico} model, we evaluate the proposed hybrid reservoir on the Mackey--Glass chaotic time-series prediction benchmark. The system achieves normalized root mean square error (NRMSE) values of approximately $0.33$--$0.40$ for prediction horizons $H=1$ to $5$, and exhibits measurable short-term memory as encoded in the distributed spatiotemporal patterns of bacteria and biochemicals. These results motivate the future exploration of non-neural hybrid cell networks for \textit{in situ} temporal signal processing towards novel biomedical applications.
\end{abstract}


\section{Introduction}
\IEEEPARstart{T}{he} Internet of Bio-Nano Things (IoBNT) envisions heterogeneous networks in which engineered micro/nanoscale devices and biological entities cooperate inside physiological environments to collectively sense, compute, and actuate \cite{Akyildiz2015IoBNT}. For such networks, the viable exchange of information is mainly conceived through \emph{Molecular Communications} (MolCom), where data is encoded in concentrations, gradients, and reaction kinetics \cite{Farsad2016SurveyMC}. However, in biological systems, communication is not an abstraction separable from computation as the MolCom channel is frequently an active medium in which biochemical interactions simultaneously transform signals into decisions, memory, and collective behavior. This intrinsic coupling suggests that future IoBNT ecosystems can be utilized not merely as communication networks, but as novel wetware computing substrates \cite{uzun2025molecular, gomez2025communicating}.

To date, efforts to harness biological computing have largely followed two distinct paths. The first involves ``top-down'' genetic engineering to mimic digital logic gates \cite{Kwok2010}. While successful for simple control tasks, this approach often scales poorly due to metabolic burden and the stochastic nature of gene expression. The second path seeks to harness the potent information processing capabilities of neural tissue. Recent seminal works such as ``Brainoware'' \cite{Cai2023} and ``DishBrain'' \cite{Kagan2022}, where \emph{in vitro} biological neurons play video games via electrophysiological interfaces, have demonstrated the power of biological substrates for complex learning tasks. However, neural tissues are inherently difficult to maintain and integrate into distributed, non-neural environments. This raises a pivotal question: can we exploit the innate dynamics of ubiquitous, non-neural biological systems, such as bacterial collectives, for complex computing tasks?

A promising framework to bridge this gap is Physical Reservoir Computing (PRC) \cite{Nakajima2020}. PRC exploits the high-dimensional, nonlinear dynamics of a physical system (the reservoir) to map time-varying inputs into a feature space where they become linearly separable. While nervous tissue and chemical reaction networks have recently been explored as reservoirs \cite{Baltussen2024}, harnessing the \textit{spatial} heterogeneity and active motility of bacteria requires a new interface strategy. Bottom-up synthetic biology provides a promising route through \emph{artificial cells (ACs)}, which are engineered compartments that can be programmed to produce, transform, and release chemical signals while remaining modular and externally controllable. A growing experimental literature has demonstrated MolCom between artificial and living cells, including bidirectional signaling via quorum-sensing molecules \cite{Lentini2014Translate, Lentini2017TwoWay}, suggesting that ACs can function as localized actuators of biological systems in complex environments.

In this letter, we propose a hybrid computing architecture that integrates these elements, employing an \textit{E. coli} population as the computing reservoir and ACs as \emph{programmable local transducers}. Unlike genetic engineering, which modifies the internal hardware of the cell, ACs modulate the \emph{biochemical context in the microenvironment}, the information channel itself. Concretely, the ACs transduce a scalar input stream into time-varying release of attractant (AHL) and repellent molecules, generating spatially heterogeneous gradients that continuously drive the bacterial population into complex spatiotemporal configurations. We implement this architecture in an \emph{in situ} agent-based model that couples bacterial motion and metabolism with reaction--diffusion fields (including decay/flow washout), such that past inputs are retained through evolving spatiotemporal patterns of bacteria collective and biochemicals. The resulting reservoir state is sampled coarsely by voxel-based measurements of key molecular fields and bacterial density.

We evaluate the temporal information processing capability of this hybrid system using the Mackey--Glass chaotic time-series prediction benchmark \cite{Mackey1977}. In our formulation, the biological substrate provides the necessary nonlinear dynamics, while learning is confined to the external readout, consistent with PRC principles. By training on the voxelated state history, we demonstrate that the coupled bacteria--molecule dynamics can accurately forecast the chaotic signal. We quantify this performance through prediction accuracy across multiple horizons and characterize the reservoir's intrinsic short-term memory using standard memory-curve analysis.

The remainder of this letter is organized as follows. Section~II introduces the system model and problem formulation. Section~III details the RC framework and the \emph{in situ} agent-based implementation. Section~IV presents prediction and memory results. Section~V discusses implications and limitations, and Section~VI concludes the letter.

\section{System Overview and Problem Formulation}
\label{sec:systemoverview}

\begin{figure*}[t]
    \centering
    \includegraphics[width=0.75\textwidth]{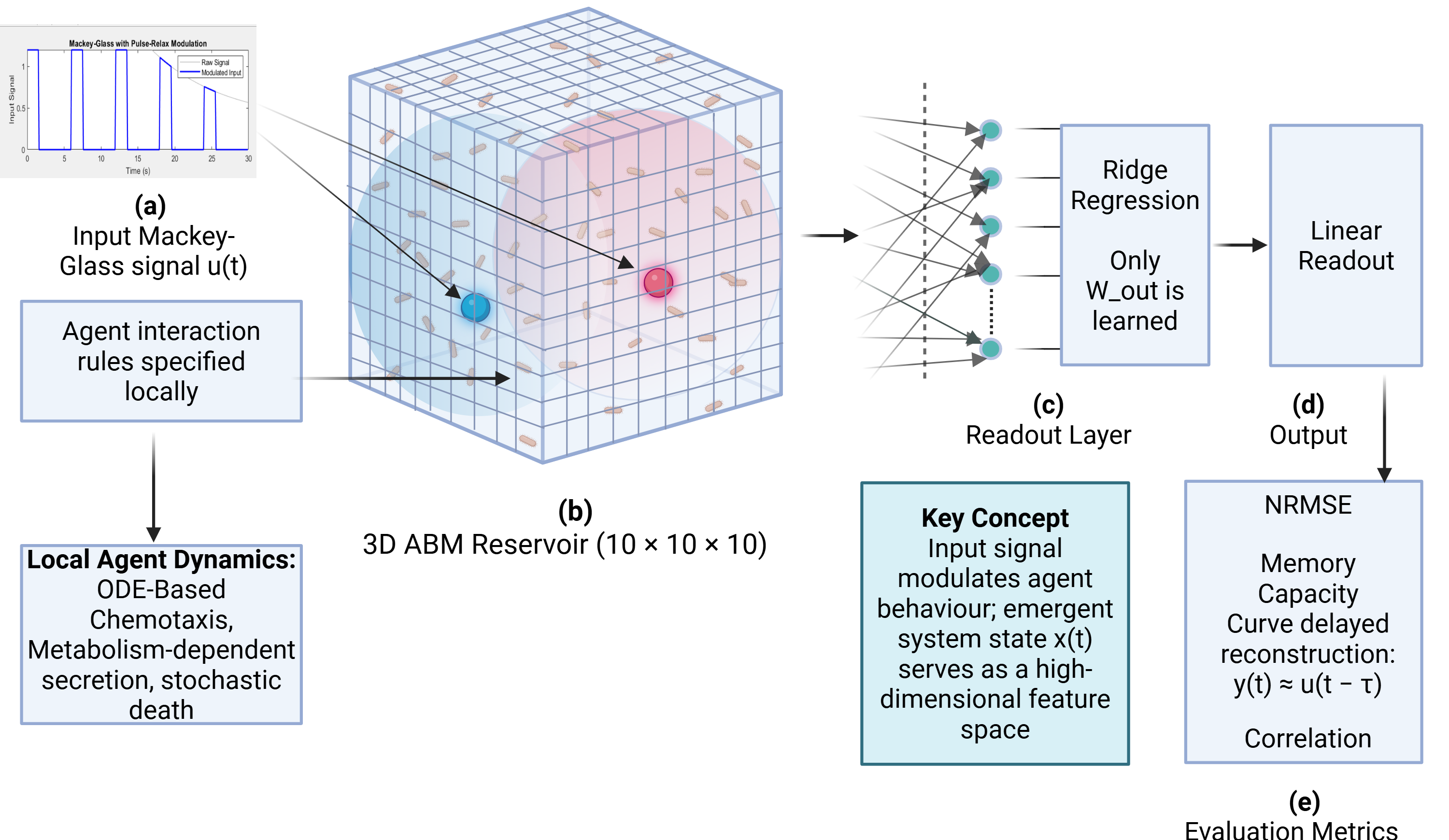}
    \caption{\textbf{Architecture of the hybrid artificial-living cell RC implemented \emph{in silico} via agent-based modeling framework.}
    \textbf{(a)} A scalar input sequence $u[n]$, derived from a chaotic time series, is encoded into time-varying chemical secretion by mobile ACs.
    \textbf{(b)} These cells function as programmable local transducers within a 3D voxelated environment, where bacteria interact through motility and metabolic dynamics in response to chemical gradients.
    \textbf{(c)} The reservoir state is captured by sampling voxel-wise molecular concentrations and bacterial density within each processing window.
    \textbf{(d)} A trained linear readout maps the biological dynamics to a predicted output.
    \textbf{(e)} Performance is quantified via NRMSE (and correlation) across prediction horizons, assessing how well the reservoir reconstructs the temporal structure of the input process.}
    \label{fig:abm_rc_arch}
\end{figure*}

We consider a hybrid artificial--living molecular reservoir implemented in an agent-based modeling (ABM) setting using the BSim framework \cite{Gorochowski2012BSim,Gorochowski2016ABM}. Computation emerges from the coupling among engineered ACs, a living bacterial population, and diffusive molecular fields (see also Fig. \ref{fig:abm_rc_arch}).

\subsection{Environment and Molecular Transport}
The environment is a 3D domain discretized into a $10 \times 10 \times 10$ voxel grid ($V=10^3$ voxels). The extracellular medium contains three key chemical species: an attractant (AHL), a repellent, and a nutrient source (glucose). The concentration $c_k(\mathbf{x},t)$ of species $k \in \{a,r,g\}$ evolves according to reaction--diffusion--advection dynamics \cite{Jamali2019}:
\begin{equation}
\frac{\partial c_k}{\partial t}
=
D_k \nabla^2 c_k
-
\nabla \cdot \left(\mathbf{v}\, c_k\right)
-
\alpha_k c_k
+
S_{\mathrm{AC},k}
+
S_{\mathrm{B},k},
\end{equation}
where $D_k$ is the diffusion coefficient, $\alpha_k$ is the decay rate, and $\mathbf{v}$ is a unidirectional flow field representing washout. The terms $S_{\mathrm{AC},k}$ and $S_{\mathrm{B},k}$ represent secretion and consumption by ACs and bacteria. The combination of diffusion, decay, and washout prevents unbounded accumulation of signals, enforcing the fading memory property essential for PRC.

\subsection{Artificial Cell Transducers}
Two mobile artificial leader cells (ACs) act as programmable input interfaces. Their role is to convert a scalar input signal $u(t)$ into time-varying secretion rates, thereby writing information into the chemical environment. The internal state of each AC is modeled as a two-stage leaky integrator:
\begin{align}
\frac{d x_{\mathrm{AC}}(t)}{dt} &= k_u u(t) - \gamma_x x_{\mathrm{AC}}(t), \\
\frac{d s_{\mathrm{AC}}(t)}{dt} &= k_x x_{\mathrm{AC}}(t) - \gamma_s s_{\mathrm{AC}}(t),
\end{align}
where $x_{\mathrm{AC}}(t)$ is a filtered internal signal and $s_{\mathrm{AC}}(t)$ represents secretion readiness. The secretion rate for molecular species $k$ is given by:
\begin{equation}
Q_{\mathrm{AC},k}(t) = \alpha_k \, s_{\mathrm{AC}}(t)\, I(t-nT_w),
\end{equation}
with the gating function $I(\tau)$ defined as:
\begin{equation}
I(\tau) =
\begin{cases}
1, & 0 < \tau < T_p, \\
0, & T_p \le \tau < T_w.
\end{cases}
\end{equation}
Here, $T_w$ is the processing window length and $T_p$ is the stimulation duration. This gating creates a relaxation phase within each window, promoting temporal mixing and partial washout of the signal.

\subsection{Bacterial Collective Dynamics}
The living \textit{E.~coli} collective constitutes the nonlinear computational core. Simulations are initialized with $N=90$ agents and evolve via growth to $N \approx 700$. Agent behavior is governed by three coupled mechanisms:

\subsubsection{Chemotaxis and Adaptation}
Bacterial sensing follows the Barkai--Leibler adaptation model \cite{Barkai1997}. Receptor methylation $m_i(t)$ for bacterium $i$ evolves as:
\begin{equation}
\frac{d m_i(t)}{dt} = k_R \bigl(1 - a_i(t)\bigr) - k_B a_i(t),
\end{equation}
where $a_i(t)$ is receptor activity and $k_R, k_B$ are rates. Motion follows run-and-tumble dynamics with tumble propensity $P_{\mathrm{tumble},i}(t) = T_0 \exp[\gamma (a_i(t)-a_0)]$ \cite{Berg2004}. Since methylation depends on past exposure, bacterial motility exhibits intrinsic fading memory.

\subsubsection{Metabolism and Growth}
Metabolic energy $E_i(t)$ follows Monod-type kinetics:
\begin{equation}
\frac{dE_i(t)}{dt} =
\frac{V_{\max} \, C_{g,i}(t)}{K_g + C_{g,i}(t)}
-
\eta E_i(t),
\end{equation}
where $C_{g,i}(t)$ is local glucose concentration, $V_{\max}$ is uptake rate, and $\eta$ is basal expenditure. Replication occurs when energy exceeds a threshold, while death results from starvation, toxicity, or stochastic baseline processes.

\subsubsection{Quorum Sensing}
Bacteria produce AHL (quorum sensing signal) in a glucose-dependent manner \cite{Gorochowski2016ABM}:
\begin{equation}
f_{\mathrm{AHL}}(G) = \alpha_{\mathrm{AHL}} \cdot \frac{G}{K_{G,\mathrm{AHL}} + G}.
\end{equation}
This resource-dependence couples the communication channel capacity to metabolic state, acting as an endogenous control on the reservoir's information processing capacity \cite{Dambre2012}.

\section{Reservoir Computing Framework}
\label{sec:PRC}

We utilize the hybrid system described above as a PRC to process temporal information.

\subsection{Input Signal and Encoding}
The system is driven by the Mackey--Glass (MG) chaotic time-series, a benchmark for temporal processing defined by:
\begin{equation}
\frac{d x_{\mathrm{MG}}(t)}{dt}
=
\beta \frac{x_{\mathrm{MG}}(t-\tau)}{1 + x_{\mathrm{MG}}(t-\tau)^n}
-
\gamma x_{\mathrm{MG}}(t),
\end{equation}
with $\beta = 0.2$, $\gamma = 0.1$, $n = 10$, and $\tau = 17$. The continuous trajectory is sampled to form a discrete sequence $\{u[n]\}$. Each sample $u[n]$ drives the ACs during processing window $n$, modulating the biochemical gradients as described in Eq.~(2)--(5).

\subsection{Reservoir State Extraction}
Within each input window $n$, the physical state of the reservoir is captured by sampling the voxel-wise distributions of attractant $\mathbf{c}_a[n]$, repellent $\mathbf{c}_r[n]$, and bacterial density $\mathbf{p}[n]$. These are concatenated to form the high-dimensional reservoir state vector:
\begin{equation}
\mathbf{r}[n]
=
\bigl[\mathbf{c}_a[n]^{\mathsf{T}},\, \mathbf{c}_r[n]^{\mathsf{T}},\, \mathbf{p}[n]^{\mathsf{T}}\bigr]^{\mathsf{T}}
\in \mathbb{R}^{N},
\end{equation}
where $N=3V=3000$. An initial wash-in period is discarded to reduce dependence on initial conditions.

\subsection{Readout and Problem Formulation}
To define the computational task, we employ a linear readout trained to estimate a target scalar $y[n]$ from the reservoir states. To expose temporal structure to the readout without altering the reservoir, we use a tapped-delay embedding of depth $k$:
\begin{equation}
\boldsymbol{\phi}[n] = [\mathbf{r}[n]^{\mathsf{T}}, \mathbf{r}[n-1]^{\mathsf{T}}, \ldots, \mathbf{r}[n-k]^{\mathsf{T}}]^{\mathsf{T}}.
\end{equation}
The readout weights $\mathbf{W}$ are computed via ridge regression to minimize the cost function:
\begin{equation}
\mathbf{W} = \arg\min_{\mathbf{W}} \|\mathbf{Y} - \mathbf{\Phi}\mathbf{W}\|^2 + \lambda\|\mathbf{W}\|^2,
\end{equation}
where $\mathbf{\Phi}$ is the stacked feature matrix, $\mathbf{Y}$ contains the targets, and $\lambda$ is a regularization parameter. The primary task is $H$-step-ahead prediction, where the target is a future value of the input sequence:
\begin{equation}
y[n] = u[n + H].
\end{equation}

\section{Experimental Evaluation Setup}
\label{sec:setup}

\subsection{Data Partitioning and Preprocessing}
\label{sec:preprocessing}
The reservoir state trajectory $\{\mathbf{r}[n]\}$ is collected sequentially. To assess generalization, the data is split into training ($70\%$), validation ($15\%$), and test ($15\%$) sets without shuffling. To mitigate sensitivity to the choice of split point, we repeat evaluations across six different temporal split offsets and report the median performance.
Before training, Principal Component Analysis (PCA) is applied to the embedded features $\boldsymbol{\phi}[n]$ to reduce dimensionality and collinearity. The PCA basis is fitted only on the training set and retains components explaining approximately $95\%$ of the variance.

\subsection{Performance Metrics}
Prediction performance is quantified on the held-out test set using the Normalized Root Mean Square Error (NRMSE) and the Pearson correlation coefficient between the predicted output $\hat{y}[n]$ and the target $y[n]$.

\subsection{Memory Capacity Analysis}
\label{sec:memorycapacity}
To characterize the intrinsic short-term memory of the reservoir independently of the specific prediction task, we compute the linear Memory Capacity (MC). This involves training independent readouts to reconstruct delayed versions of the input $u[n-d]$ from the current state $\mathbf{r}[n]$. The capacity is defined as:
\begin{equation}
\mathrm{MC} = \sum_{d=1}^{d_{\max}} R^2(d),
\end{equation}
where $R^2(d)$ is the coefficient of determination for delay $d$.

\section{Results}

We report prediction and memory results for the hybrid ABM reservoir described in Sections \ref{sec:systemoverview}--\ref{sec:setup}. Unless otherwise stated, performance statistics correspond to the median across the six temporal split offsets described in Section \ref{sec:preprocessing}.

\subsection{Prediction Accuracy Across Horizons}
\begin{figure}[t]
    \centering
    \includegraphics[width=0.9\columnwidth]{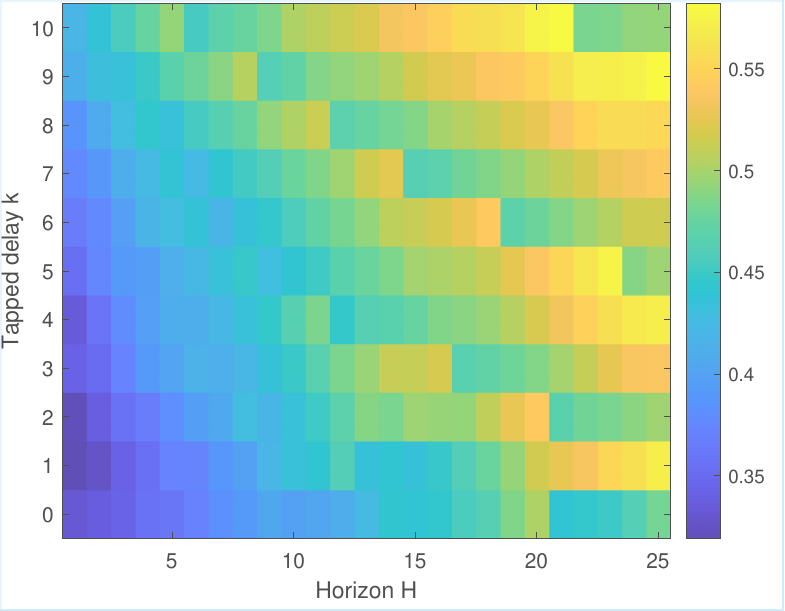}
    \caption {\textbf{NRMSE heatmap across prediction horizon ($H$) and tapped-delay depth ($k$), evaluated over multiple temporal splits.} Lowest error is achieved at short horizons, with gradually increasing error for larger $H$, consistent with fading memory and limited long-range predictability of the hybrid molecular reservoir.}
    \label{fig:heat_map}
\end{figure}

Fig.~\ref{fig:heat_map} summarizes forecasting accuracy as a function of prediction horizon $H$ and tapped-delay depth $k$. The error increases smoothly with $H$, indicating a gradual loss of predictability as the task becomes more dependent on longer input histories. For short horizons ($H=1$--$5$), the median NRMSE remains low (approximately $0.33$--$0.40$), while longer-horizon forecasts exhibit progressively higher error. The absence of abrupt performance collapses or irregular ``spikes'' across $H$ suggests that the observed degradation is driven primarily by finite reservoir memory and the inherent difficulty of long-range prediction for chaotic dynamics, rather than numerical instability or overfitting.

\subsection{Effect of Tapped-Delay Depth}

\begin{figure}[t]
    \centering
    \includegraphics[width=0.9\columnwidth]{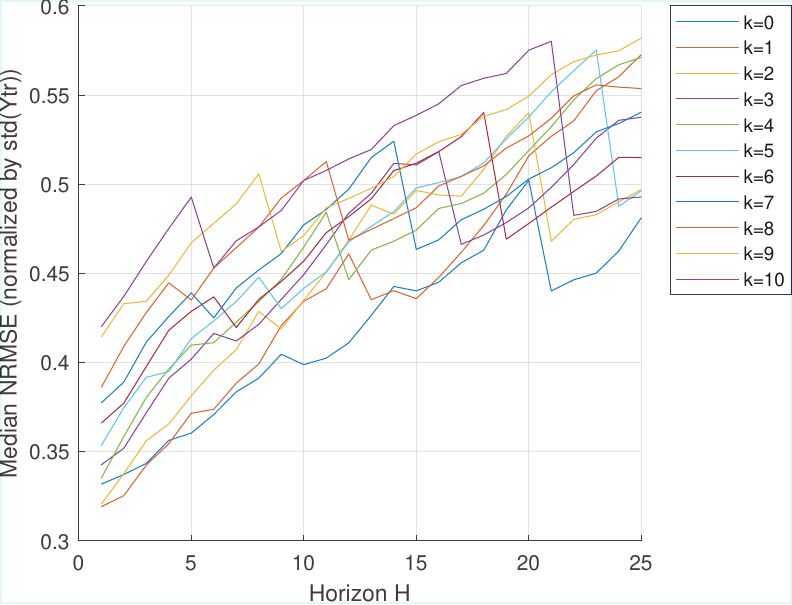}
    \caption {\textbf{Median NRMSE versus prediction horizon ($H$) for varying tapped-delay depths ($k$).} Temporal embedding ($k>0$) improves prediction, especially at longer horizons, indicating that task-relevant information is distributed across recent reservoir states.}
    \label{fig:k_values}
\end{figure}

Temporal embedding improves prediction accuracy across most horizons. As shown in Fig.~\ref{fig:k_values}, readouts with nonzero delay depth ($k>0$) consistently outperform the memoryless configuration ($k=0$), with the largest gains appearing at intermediate and longer horizons where task-relevant information is more distributed across time. Moderate depths ($k\approx 3$--$6$) provide noticeable improvements at short-to-medium horizons, whereas larger depths (e.g., $k=10$) tend to be more beneficial when forecasting further ahead. This trend is consistent with the interpretation that increasing $k$ exposes more of the reservoir's recent trajectory to the linear readout, allowing it to better leverage the reservoir's intrinsic fading memory without modifying the reservoir dynamics.

\subsection{Correlation Behavior}

\begin{figure}[t]
    \centering
    \includegraphics[width=0.9\columnwidth]{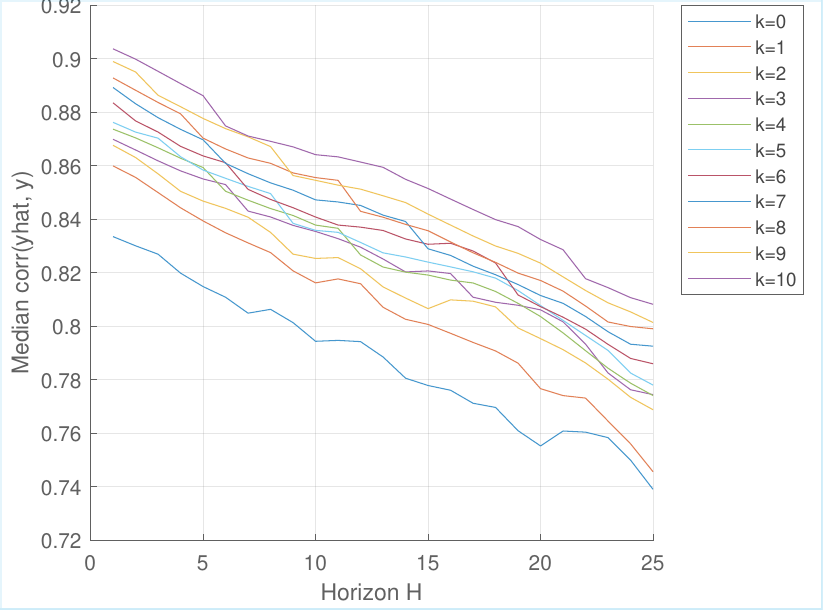}
     \caption {\textbf {Median correlation between predicted and target signals versus prediction horizon ($H$).} Correlation degrades gradually with $H$, with larger tapped-delay depths maintaining superior long-horizon performance.}
    \label{fig:median_correlation}
\end{figure}

The correlation between predictions $\hat{y}[n]$ and targets $y[n]=u[n+H]$ provides a complementary view of performance. Fig.~\ref{fig:median_correlation} shows that correlation decreases monotonically with increasing $H$ while remaining positive throughout, reflecting a progressive loss of alignment rather than a qualitative failure mode. Higher tapped-delay depths maintain systematically higher correlation at larger horizons, corroborating the NRMSE trends and confirming that temporal embedding improves the readout's ability to extract predictive structure from the reservoir trajectory.

\subsection{Memory Curve, Memory Capacity, and Forecasting Limits}

\begin{figure}[t]
    \centering
    \includegraphics[width=0.9\columnwidth]{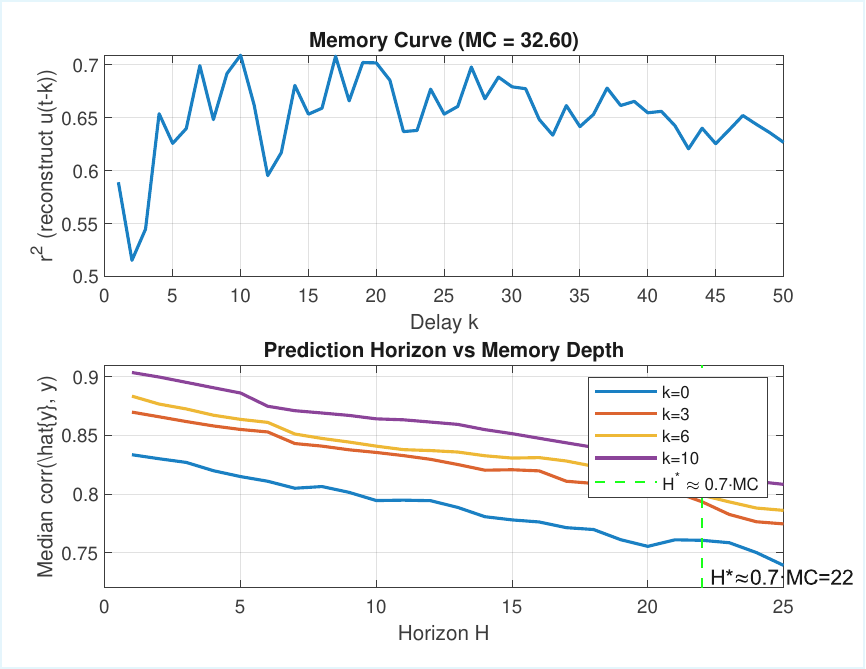}
     \caption {\textbf{Linking intrinsic memory to forecasting.} (Top) Memory curve $R^2(d)$ obtained by reconstructing delayed inputs $u[n-d]$ from the instantaneous reservoir state $\mathbf{r}[n]$, yielding $\mathrm{MC}\approx 32.6$. (Bottom) Median prediction correlation versus horizon $H$ for selected tapped-delay depths. Performance degradation becomes pronounced near $H^\star \approx 0.7\,\mathrm{MC}$, providing an internal consistency check between intrinsic reservoir memory and achievable prediction horizon.}
    \label{fig:memory_curve}
\end{figure}

To quantify intrinsic short-term memory independently of multi-step forecasting, we reconstruct delayed copies of the input from the instantaneous reservoir state $\mathbf{r}[n]$ and compute a memory curve $R^2(d)$ as described in Section \ref{sec:memorycapacity}, where $d$ denotes the reconstruction delay (in windows). The resulting curve in Fig.~\ref{fig:memory_curve} yields a total linear memory capacity of $\mathrm{MC}\approx 32.6$. Beyond the aggregate value, the non-monotonic structure of the curve indicates that memory is not governed by a single diffusion time constant; instead, it is distributed across multiple interacting processes (chemotactic adaptation, quorum-mediated coupling, growth/death, and washout), which together create a richer set of effective temporal scales.

Importantly, the independently measured $\mathrm{MC}$ provides a quantitative explanation for the observed long-horizon prediction limits. In our experiments, prediction quality degrades markedly around $H\approx 20$--$23$, which is consistent with $H^\star \approx 0.7\,\mathrm{MC}$. This alignment supports the interpretation that long-range forecasting is primarily \emph{memory-limited} rather than constrained by readout expressivity.

Taken together, Figs.~\ref{fig:heat_map}--\ref{fig:memory_curve} demonstrate that the hybrid AC--bacteria system realizes a stable fading-memory physical reservoir in which information about past inputs is retained and transformed in distributed spatiotemporal patterns of bacteria and molecular fields. 

\section{Conclusion}
We presented an \emph{in silico} hybrid artificial--living RC architecture in which mobile artificial cells encode an external time-series into localized secretion of attractant and repellent molecules, and a living \textit{E.~coli} collective provides the nonlinear spatiotemporal dynamics required for temporal information processing. Using a voxel-sampled state representation and a ridge-regression readout with tapped-delay embedding, the system achieves strong short-horizon forecasting on the Mackey--Glass benchmark and degrades smoothly with prediction horizon, consistent with fading-memory dynamics. An independent memory-capacity analysis yields $\mathrm{MC}\approx 32.6$ and predicts the onset of long-horizon performance degradation, indicating that forecasting is fundamentally constrained by intrinsic reservoir memory rather than readout complexity. Overall, the results support the view that hybrid cell collectives can serve as programmable wetware reservoirs in which molecular transport and biophysical coupling jointly realize communication and computation. Future work will focus on experimental realizations and task generalization to application-driven \emph{in situ} biochemical signal processing.

\bibliographystyle{IEEEtran}
\bibliography{references}

@article{Akyildiz2015IoBNT,
  author  = {Akyildiz, Ian F. and Pierobon, Massimiliano and Balasubramaniam, Sasitharan and Koucheryavy, Yevgeni},
  title   = {The {Internet} of {Bio}-{Nano} Things},
  journal = {IEEE Commun. Mag.},
  volume  = {53},
  number  = {3},
  pages   = {32--40},
  year    = {2015},
  month   = mar,
  doi     = {10.1109/MCOM.2015.7060516}
}

@article{Farsad2016SurveyMC,
  author  = {Farsad, Nariman and Yilmaz, H. Birkan and Eckford, Andrew and Chae, Chan-Byoung and Guo, Weisi},
  title   = {A Comprehensive Survey of Recent Advancements in Molecular Communication},
  journal = {IEEE Commun. Surveys Tuts.},
  volume  = {18},
  number  = {3},
  pages   = {1887--1919},
  year    = {2016},
  doi     = {10.1109/COMST.2016.2527741}
}

@article{Kwok2010,
  author  = {Kwok, Roberta},
  title   = {Five Hard Truths for Synthetic Biology},
  journal = {Nature},
  volume  = {463},
  number  = {7279},
  pages   = {288--290},
  year    = {2010},
  month   = jan,
  doi     = {10.1038/463288a}
}

@article{Cai2023,
  author  = {Cai, Hongwei and Ao, Zheng and Tian, Chunhui and Wu, Zhuhao and Liu, Hongcheng and Tchieu, Jason and Gu, Mingxia and Mackie, Ken and Guo, Feng},
  title   = {Brain Organoid Reservoir Computing for Artificial Intelligence},
  journal = {Nat. Electron.},
  volume  = {6},
  number  = {12},
  pages   = {1032--1039},
  year    = {2023},
  month   = dec,
  doi     = {10.1038/s41928-023-01069-w}
}

@article{Kagan2022,
  author  = {Kagan, Brett J. and Kitchen, Andy C. and Tran, Nhi T. and Habibollahi, Forough and Khajehnejad, Moein and Parker, Bradyn J. and Bhat, Anjali and Rollo, Ben and Razi, Adeel and Friston, Karl J.},
  title   = {In Vitro Neurons Learn and Exhibit Sentience When Embodied in a Simulated Game-World},
  journal = {Neuron},
  volume  = {110},
  number  = {23},
  pages   = {3952--3969.e8},
  year    = {2022},
  month   = dec,
  doi     = {10.1016/j.neuron.2022.09.001}
}

@article{Nakajima2020,
  author  = {Nakajima, Kohei},
  title   = {Physical Reservoir Computing---An Introductory Perspective},
  journal = {Jpn. J. Appl. Phys.},
  volume  = {59},
  number  = {6},
  pages   = {060501},
  year    = {2020},
  doi     = {10.35848/1347-4065/ab8d4f}
}

@article{Baltussen2024,
  author  = {Baltussen, Mathieu G. and de Jong, Thijs J. and Duez, Quentin C. N. and Robinson, William E. and Huck, Wilhelm T. S.},
  title   = {Chemical Reservoir Computation in a Self-Organizing Reaction Network},
  journal = {Nature},
  volume  = {631},
  number  = {8021},
  pages   = {549--555},
  year    = {2024},
  month   = jul,
  doi     = {10.1038/s41586-024-07567-x}
}

@article{Lentini2014Translate,
  author  = {Lentini, Roberta and Santero, Silvia Perez and Chizzolini, Fabio and Cecchi, Dario and Fontana, Jason and Marchioretto, Marta and Del Bianco, Cristina and Terrell, Jessica L. and Spencer, Amy C. and Martini, Laura and Forlin, Michele and Assfalg, Michael and Dalla Serra, Mauro and Bentley, William E. and Mansy, Sheref S.},
  title   = {Integrating Artificial with Natural Cells to Translate Chemical Messages That Direct {E}.~coli Behaviour},
  journal = {Nat. Commun.},
  volume  = {5},
  pages   = {4012},
  year    = {2014},
  month   = may,
  doi     = {10.1038/ncomms5012}
}

@article{Lentini2017TwoWay,
  author  = {Lentini, Roberta and Mart{\'\i}n, Nicol{\'a}s Y. and Forlin, Michele and Belmonte, Lucio and Fontana, Jason and Cornella, Miquel and Martini, Laura and Tamburini, Sandra and Bentley, William E. and Mansy, Sheref S.},
  title   = {Two-Way Chemical Communication between Artificial and Natural Cells},
  journal = {ACS Cent. Sci.},
  volume  = {3},
  number  = {2},
  pages   = {117--123},
  year    = {2017},
  month   = feb,
  doi     = {10.1021/acscentsci.6b00330}
}

@article{Mackey1977,
  author  = {Mackey, Michael C. and Glass, Leon},
  title   = {Oscillation and Chaos in Physiological Control Systems},
  journal = {Science},
  volume  = {197},
  number  = {4300},
  pages   = {287--289},
  year    = {1977},
  month   = jul,
  doi     = {10.1126/science.267326}
}

@article{Gorochowski2012BSim,
  author  = {Gorochowski, Thomas E. and Matyjaszkiewicz, Antoni W. and Todd, Thomas and Oak, Neeraj and Kowalska, Kira and Reid, Stephen and Tsaneva-Atanasova, Krasimira T. and Savery, Nigel J. and Grierson, Claire S. and di Bernardo, Mario},
  title   = {{BSim}: An Agent-Based Tool for Modeling Bacterial Populations in Systems and Synthetic Biology},
  journal = {PLOS One},
  volume  = {7},
  number  = {8},
  pages   = {e42790},
  year    = {2012},
  month   = aug,
  doi     = {10.1371/journal.pone.0042790}
}

@article{Gorochowski2016ABM,
  author  = {Gorochowski, Thomas E.},
  title   = {Agent-Based Modelling in Synthetic Biology},
  journal = {Essays Biochem.},
  volume  = {60},
  number  = {4},
  pages   = {325--336},
  year    = {2016},
  month   = nov,
  doi     = {10.1042/EBC20160037}
}

@article{Jamali2019,
  author  = {Jamali, Vahid and Ahmadzadeh, Arman and Wicke, Wayan and Noel, Adam and Schober, Robert},
  title   = {Channel Modeling for Diffusive Molecular Communication---A Tutorial Review},
  journal = {Proc. IEEE},
  volume  = {107},
  number  = {7},
  pages   = {1256--1301},
  year    = {2019},
  month   = jul,
  doi     = {10.1109/JPROC.2019.2919455}
}

@article{Barkai1997,
  author  = {Barkai, Naama and Leibler, Stanislas},
  title   = {Robustness in Simple Biochemical Networks},
  journal = {Nature},
  volume  = {387},
  number  = {6636},
  pages   = {913--917},
  year    = {1997},
  month   = jun,
  doi     = {10.1038/43199}
}

@book{Berg2004,
  author    = {Berg, Howard C.},
  title     = {{E}.~coli in Motion},
  publisher = {Springer},
  address   = {New York, NY, USA},
  year      = {2004},
  doi       = {10.1007/b97370}
}

@article{Dambre2012,
  author  = {Dambre, Joni and Verstraeten, David and Schrauwen, Benjamin and Massar, Serge},
  title   = {Information Processing Capacity of Dynamical Systems},
  journal = {Sci. Rep.},
  volume  = {2},
  pages   = {514},
  year    = {2012},
  doi     = {10.1038/srep00514}
}

@article{uzun2025molecular,
  title={Molecular Communication Channel as a Physical Reservoir Computer},
  author={Uzun, Mustafa and Ikiz, Kaan Burak and Kuscu, Murat},
  journal={arXiv preprint arXiv:2504.17022},
  year={2025}
}

@article{gomez2025communicating,
  title={Communicating Smartly in Molecular Communication Environments: Neural Networks in the Internet of Bio-Nano Things},
  author={G{\'o}mez, Jorge Torres and Hofmann, Pit and Debus, Lisa Y and Ba{\c{s}}aran, Osman Tugay and Lotter, Sebastian and Khanzadeh, Roya and Angerbauer, Stefan and Unluturk, Bige Deniz and Abadal, Sergi and Haselmayr, Werner and others},
  journal={arXiv preprint arXiv:2506.20589},
  year={2025}
}

\end{document}